\documentclass[preprint2]{aastex}
\usepackage{natbib}

\def\apjs{ApJS}

\def\apjl{ApJ}
\def\aap{A\&A}

\def\asec{\ifmmode ^{\prime\prime}\else$^{\prime\prime}$\fi}

\def\it{\sl}
\def\degs{\ifmmode ^{\circ}\else$^{\circ}$\fi}
\def\amin{\ifmmode ^{\prime}\else$^{\prime}$\fi}
\def\asec{\ifmmode ^{\prime\prime}\else$^{\prime\prime}$\fi}
\def\fm{\hbox{$.\!\!^{\rm m}$}}            
\def\fdg{\hbox{$.\!\!^\circ$}}          
\def\farcs{\hbox{$.\!\!^{\prime\prime}$}}  

\def\degs{\ifmmode ^{\circ}\else$^{\circ}$\fi}
\def\amin{\ifmmode ^{\prime}\else$^{\prime}$\fi}
\def\farcm{\hbox{$.\mkern-4mu^\prime$}}
\def\eqalign#1{\null\,\vcenter{\openup1\jot \m@th
   \ialign{\strut\hfil$\displaystyle{##}$&$\displaystyle{{}##}$\hfil
   \crcr#1\crcr}}\,}

\slugcomment{to appear in Astrophysical J. Letters, 45.}

\shorttitle{The Vela pulsar and its likely counter-jet in the $K_s$- band}
\shortauthors{Zyuzin et al.}

\begin{document}


\title{The Vela pulsar and its likely counter-jet in the $K_s$ band 
    }


\author{D.~Zyuzin,  Yu.~Shibanov\altaffilmark{1} and A.~Danilenko}
\affil{Ioffe Physical Technical Institute, Politekhnicheskaya
 26, St. Petersburg, 194021, Russia}

\author{R.~E.~Mennickent}
\affil{Department of Astronomy, Universidad de Concepcion, Casilla 160-C, Concepcion, Chile}


\and

\author{S.~Zharikov}
\affil{Observatorio Astron\'{o}mico Nacional SPM, Instituto de
 Astronom\'{i}a, UNAM, Ensenada, BC, Mexico}


\altaffiltext{1}{St.~Petersburg State Polytechnical Univ., Politekhnicheskaya 29, St. Petersburg, 195251, Russia}


\begin{abstract}
 We report the first  high spatial resolution near-infrared imaging of the 
 Vela pulsar in the $K_s$ band 
 obtained with the new  adaptive optics system 
 recently mounted on the Gemini-South telescope. 
 For the first time, we have firmly detected the pulsar 
 in this band with $K_s$~$\approx$ 21\fm8, and have resolved in detail an extended  
 feature barely detected 
 previously in the immediate vicinity of the  pulsar in the $J_sH$ bands.  
 The  pulsar $K_s$ flux  is fully consistent with  the
 extension of the flat optical spectrum of the pulsar towards the infrared  
 and  does not confirm the strong infrared flux excess in the pulsar emission suggested 
 earlier  by the low spatial resolution data.  
 The extended feature is about two times 
 brighter than the pulsar and is likely associated 
 with its X-ray counter-jet. 
 It extends $\sim$~2\asec~southwards of the pulsar along the X-ray counter-jet   
 and shows knot-like structures and a red spectrum.    
\end{abstract}


\keywords{
infrared: stars --- stars: neutron --- pulsars: individual (the Vela pulsar)
}



\section{Introduction}

After the young Crab  and 
B0540$-$69  pulsars with the visual magnitudes of 16\fm5 and 22\fm6 
the 11 kyr old Vela pulsar with the magnitude of 23\fm6 is the third brightest in the optical   
among all isolated neutron stars (NSs) known. 
The relative brightness has allowed to perform detailed optical studies  
including the successful timing, spectral, and polarization  observations.   
Similar to the Crab pulsar, Vela has an almost flat spectrum 
from the near infrared (IR) to the ultra violet (UV) \citep{mignani2007AsAp}. 
The similarity has been unexpectedly  
broken by recent \textit{Spitzer} observations in the mid-IR, where the Vela pulsar 
has shown a strong flux excess over its optical-UV spectrum extension 
towards IR \citep{danilenko2011MNRAS}. This is completely different from 
Crab whose mid-IR and optical-UV spectra are described by a single power law 
\citep{sandberg2009AsAp}. 
Similar excesses have been detected only for two  magnetars 4U 0142+61 
and 1E 2259+586  \citep{wang2006Natur,kaplan2009ApJ}, where they were interpreted 
as an emission from hypothetical fall-back X-ray irradiated discs around those NSs. 
However, Vela is not a magnetar. 
It is an ordinary rotation powered pulsar emitting 
from radio to gamma-rays and powering, like Crab,  a bright (in X-rays) torus-like pulsar wind 
nebula (PWN) with polar jets \citep{helfand2001ApJ}. The fall-back disk survival around such 
an active pulsar with a  strong relativistic particle wind appears to be problematic \citep[see, e.g.,][]{jones2007MNRAS}.   
Two other possibilities have been suggested to  
explain the excess \citep{danilenko2011MNRAS}: a complicated distribution  function of emitting particles  
in the NS magnetosphere;   a possible contamination of the pulsar flux by  
an unresolved PWN structure.  
The first one looks very unusual, while the second  
is supported by the presence of a faint  nebulosity 2\asec~away from the pulsar tentatively detected in  the 
near-IR VLT/ISAAC $J_s$ and  $H$ images  \citep{shibanov2003AsAp}. The images had a higher spatial 
resolution than the \textit{Spitzer} ones. 
It is projected onto the origin of the pulsar X-ray counter-jet 
and has a  red colour, which could, in principle, explain the mid-IR excess \citep{danilenko2011MNRAS}. 
To check that, one needs a higher spatial resolution imaging in the near-IR. 

The Vela pulsar has never been observed in the $K$ band. Motivated by this and by the excess 
problem, we have carried out a high spatial resolution imaging of the Vela field in the 
$K_s$ band with the new generation of 
the adaptive optics (AO) system  
recently mounted on the Gemini-South telescope \citep{carrasco2012SPIE}.  The observations and data reduction 
are described in Sect. 2, the results are presented in Sect. 3 and discussed in Sect. 4.

\section{Gemini-South data}

\subsection{Observations, data reduction, and calibration}
The Vela pulsar  was observed on January 30, 2013 
in the  $K_s$ band  with the  Gemini Multi-Conjugate Adaptive
Optics System (GeMS) and  its near-infrared imager, the Gemini South Adaptive Optics Imager (GSAOI),  
mounted on the Gemini-South telescope\footnote{http://www.gemini.edu/sciops/instruments/gsaoi/documents \\
http://www.gemini.edu/sciops/instruments/gems/documents}. The observations were carried out 
in the service mode during 
the GeMS + GSAOI System Verification science program.
The GSAOI science array is a 2$\times$2 mosaic  of four Rockwell HAWAII-2RG detectors forming  
a 4080~$\times$~4080 pixel focal plane
with a field of view  of $85\asec \times 85\asec$ and a pixel scale of 0\farcs02. 
Each detector also contains a programmable On-Detector Guide Window (ODGW) which can 
provide a tip-tilt information for a combination of up to four natural guide stars  in GeMS. 
Three  natural optical guide  stars, NOMAD 0448-0138807 ($R$~$\sim$ 15\fm6), 
0448-0138766 (14\fm2), and 0448-0138794 (14\fm3) were used
for the  CANOPUS  tip-tilt wave front sensor (CWFS), which is the AO bench of GeMS. 
The latter star was also the
ODGW  infrared guide star. 
The pulsar was exposed on chip 1 of the GSAOI array and we focus below on the data 
obtained  from this chip.
The observing conditions were photometric with  
seeing $\la$~0\farcs55.

We have obtained 19 dithered 100 s science exposures with $\sim$10\arcsec~offsets at an airmass of $\sim$ 1.04. 
Gemini GeMS + GSAOI twilight flat-fields
were taken and
used to create a master flat-field frame.
Data reduction, linearity correction, sky subtraction, flat-fielding
and bad pixel correction were performed using the \texttt{IRAF} \texttt{gemini.gsaoi} tool.
The effective mean seeing, defined as the full width at half maximum (FWHM) of a stellar profile, 
on the  chip 1 AO corrected frames was  in the range of 0\farcs07--0\farcs09, depending on the individual exposure and 
a source position on the frame.  The best   
seeing concentrated within $\sim$0\farcm3 region around the pulsar, in accordance with expectations for our AO observational setup.  
Differences of instrumental magnitudes of several field stars around the pulsar in different exposures  were less than 1\%. 
The reduced frames were aligned to a single reference frame of the
best image quality and summed\footnote{final image located on \\ http://www.ioffe.ru/astro/NSG/obs/index.html}.
The resulting effective mean seeing 
and  integration time were 0\farcs085 and 1900 s, respectively. 
The  magnitude zero-point of $C_{K_s}$~= 25\fm53(4) (instrumental fluxes in ADUs) for the chip 1 was derived based 
on the photometric standards  (9132, 9136, 9144, 9146) 
from \citet{persson1998AJ} obtained   
on the same night. The atmospheric extinction of 0\fm065
in $K_s$ was taken from the GSAOI web-site.  
\begin{figure*}[t]
  \setlength{\unitlength}{1mm}
  \begin{center}
    \begin{picture}(145,52)(0,0)
      \put (-10,0) {\includegraphics[scale=0.43]{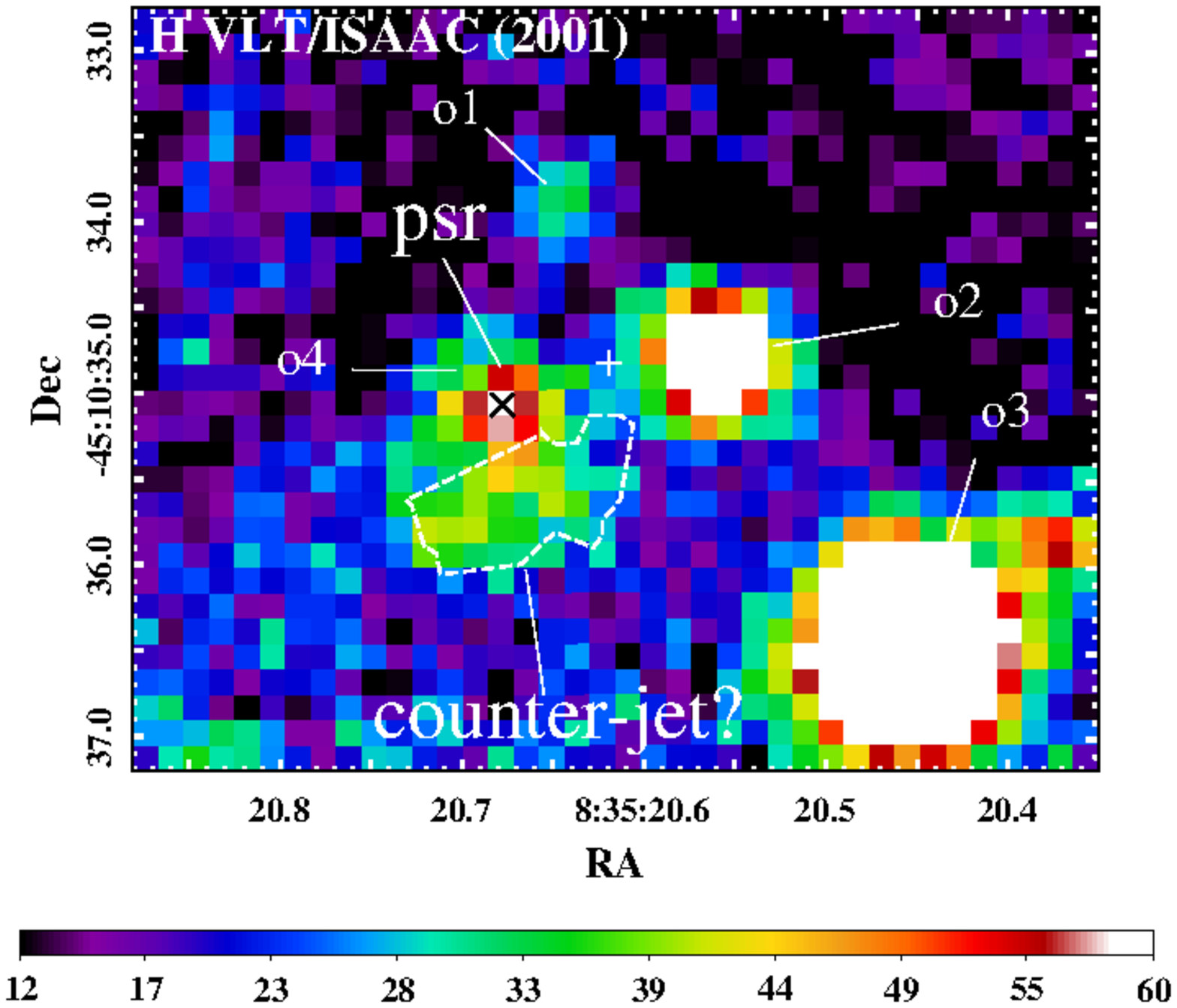}}
    \put (74,0) {\includegraphics[scale=0.43]{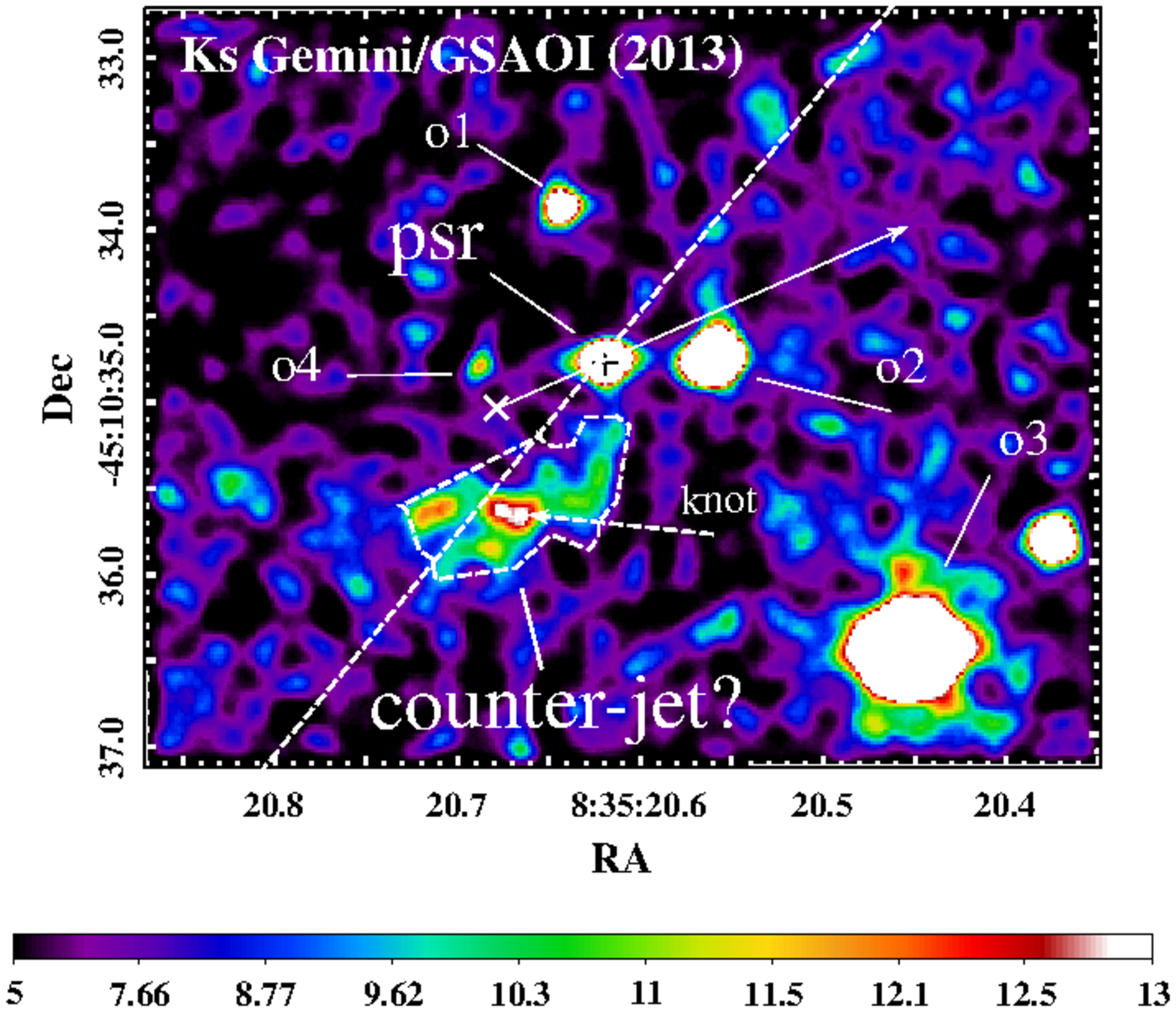}} 
    \end{picture}
  \end{center}
  \caption{
  5\farcs5~$\times$~4\farcs5~fragment 
  of the Vela pulsar vicinity as seen with the VLT/ISAAC ({\sl left}) and Gemini/(GeMS + GSAOI) ({\sl right})
  in the $H$   and $K_s$ 
  bands. The pulsar and  
  nearby field objects are marked using  notations  
  from \citet{shibanov2003AsAp}, except ``o4'', which is detected only with Gemini. 
  The $K_s$ image is smoothed with 
   a seven-pixel Gaussian kernel   
  to better underline the morphology of the extended feature, 
  presumably the counter-jet emanating from the pulsar. 
  The long-dashed line in the right frame
  indicates the 
  projection  of the X-ray counter-jet axis on the sky. 
  Symbols ``$\times$'' and ``+''  
  mark the pulsar positions  at  the epochs of the VLT 
  (2001) and Gemini (2013) observations, respectively.
  The long and short arrows indicate the direction of the pulsar proper motion 
  and the position of a bright knot within the extended feature, respectively. 
  The polygon is  
  the aperture used for  photometry of the 
  feature. 
  }
  \label{fig:gsao-im}
\end{figure*}
\subsection{Astrometry}
We have performed the relative astrometry of the  $K_s$-band resulting image
using the VLT/ISAAC $J_s$-band  image obtained
twelve years ago
by \citet{shibanov2003AsAp} as a reference frame. 
Nine unsaturated isolated stars located in the pulsar vicinity were selected as
the reference points, and used in the \texttt{IRAF} \texttt{geomap/ccmap} tools 
to obtain the plate solutions.
Formal  {\sl rms} uncertainties of the astrometric fit were   
$\la$~0\farcs014 with maximum residuals of $\la$~0\farcs032 for both coordinates.
Taking into account the uncertainties of the reference star positions, which were  
$\la$~8 mas and $\la$~0.2 mas 
for the VLT and Gemini images, respectively, 
a conservative 1$\sigma$  
referencing uncertainty of the Gemini image with respect to the VLT ones  
is $\la$~0\farcs02. Adopting the VLT  absolute astrometric image referencing from \citet{shibanov2003AsAp},   
the absolute astrometric uncertainties 
of the Gemini $K_s$-band  image
is $\approx$~0\farcs21. 


\section{Results}
\subsection{Identification of the pulsar and a likely counterpart of its counter-jet}
A fragment of the resulting $K_s$ image  
demonstrating the Vela pulsar vicinity is shown in the {\sl right panel} of Fig.~\ref{fig:gsao-im}.
It is compared with a similar fragment obtained in the $H$ band with the VLT/ISAAC by \citet{shibanov2003AsAp} ({\sl left panel}).
We consider a point-like object marked by ``+'' and detected in $K_s$ at $\sim$~20$\sigma$ significance 
as the Vela pulsar counterpart candidate. In the $K_s$ image, there is also an extended  feature 
adjacent to the counterpart candidate from the south. 
The feature is also barely resolved 
in the $H$ band, which means that it is not
an artifact.

The position of the suggested pulsar counterpart  in the $K_s$ image is 
shifted by 0\farcs68~$\pm$ 0\farcs05 with respect 
to the pulsar position at the epoch of the VLT observations (2001).
The shift implies a proper motion  $\mu$~= 56~$\pm$ 4~mas~yr$^{-1}$ with 
positional angle PA~= 296\fdg0~$\pm$ 4\fdg3, 
which is 
consistent  with the pulsar proper 
motion $\mu$~= 58~$\pm$ 0.1~mas~yr$^{-1}$ and PA~=  
301\fdg0~$\pm$ 1\fdg8 based on radio observations \citep{dodson2003ApJ}.
This is a strong evidence that the point-like 
object detected in the $K_s$ image is the pulsar.
None other point-like object in the pulsar vicinity  
demonstrates any significant proper motion.
     
The extended feature  
was named ``counter-jet?'' by \citet{shibanov2003AsAp}, assuming its possible association with 
the  counter-jet of the Vela  torus-like X-ray PWN. 
The Gemini AO observations  confirm this structure 
with a higher significance   
and reveal important details of its morphology. It extends southwards from the pulsar by about 
2\asec~along the X-ray counter-jet axis 
with PA~$\approx$ 130\degs~\citep{helfand2001ApJ}
marked by the dashed line in the {\sl right} panel of Fig.~\ref{fig:gsao-im}. 
This supports the association of the extended feature with the Vela X-ray counter-jet.  
In X-rays the counter-jet extends to a much larger distance  
up to  1\amin. In the near-IR it is likely that we see 
only its origin near the pulsar. 
The $K_s$-band counter-jet demonstrates a non-uniform morphology with several knots. 
The brightest knot is in the center of the feature. 
It is marked in Fig.~\ref{fig:gsao-im} and is reminiscent 
of the knot structure 0\farcs6 away from the Crab pulsar, which is also projected onto the counter-jet origin  
of the Crab PWN \citep{hester1995ApJ}.
The Vela IR counter-jet is not detected in the optical range 
with the \textit{HST} \citep{shibanov2003AsAp} whose spatial resolution is comparable to 
that of Gemini/GSAOI. This implies that it has a very red spectrum. 
Also it becomes evident, that the putative counter-jet and the pulsar cannot be resolved from each other  
in the significantly lower spatial resolution \textit{Spitzer} mid-IR images  
analyzed by \citet{danilenko2011MNRAS}.

In the  Gemini image we do not find  any other extended structure that could be identified with other parts of 
the Vela  PWN. In particular, we do not see the inner arc whose marginal detection in $J_s$ was discussed by \citet{shibanov2003AsAp}.
In addition to the  o1--o3 point-like objects from the pulsar neighborhood considered  earlier  
\citep{shibanov2003AsAp,danilenko2011MNRAS}  we find a red  point-like  source 
``o4'' detected east of the pulsar at $\sim$5$\sigma$ confidence  only in the $K_s$ image. 
It is significantly fainter than the pulsar and  
was partially overlapped 
with it  at  the VLT   
observation epoch.    
\begin{table*}[tbh]
\begin{center}
\caption{Magnitudes and fluxes of the pulsar and the counter-jet feature  
(defined by the polygon in Fig.~\ref{fig:gsao-im})
observed and dereddened with $E_{B-V}$~= 0.055(5) and $R_V$~= 3.1.}
\begin{tabular}{lllll}
\tableline\tableline
$\lambda_{eff}$            &    \multicolumn{2}{c}{Observed}     &     \multicolumn{2}{c}{Dereddened}       \\
      (band),                             & Mag.$^a$ & $\log$ F$_\nu$, & Mag. &  $\log$ F$_\nu$, \\ 
     $\mu$m                                      &      & $\mu$Jy       &      &  $\mu$Jy              \\ \hline    
    \multicolumn{5}{c}{pulsar}  \\ 
    \hline 
    1.23($J_s$)$^b$             & 22.71(10)  & 0.14(4)  & 22.66(10) & 0.16(4)   \\
    1.65($H$)$^b$               & 22.04(18)  & 0.21(7)  & 22.00(18) & 0.23(7)   \\
    \hline
    {\bf 2.16($K_s$)}           & {\bf 21.76(6)}  & {\bf 0.12(2)}  & {\bf 21.74(6)} & {\bf 0.13(2)}   \\
    \hline
    \multicolumn{5}{c}{counter-jet}  \\ 
    \hline   
    1.23($J_s$)            &  24.14(37)  & $-$0.43(15)        & 24.09(37) & $-$0.41(15)          \\
    1.65($H$)               &  22.35(10)  &  0.09(8)           & 22.31(10) & 0.11(8)  \\
    \hline
    {\bf 2.16($K_s$)}           & {\bf 21.16(8)} & {\bf 0.36(3)}         & {\bf 21.14(8)} & {\bf 0.37(3)}     \\
    \hline
    \multicolumn{5}{c}{\textit{Spitzer} data on the pulsar}  \\ 
    \multicolumn{5}{c}{+ counter-jet feature}  \\ 
    \hline
    3.6$^c$                     & 18.48(22)  & 1.06(9)             & 18.48(22)  & 1.06(9)   \\
    4.5$^c$                     & $\ga$ 16.84  & $\la$ 1.52          & $\ga$ 16.84  & $\la$ 1.52         \\
    5.8$^c$                     & 16.38(27)  & 1.51(11)             & 16.38(27) & 1.51(11)   \\
    8.0$^c$                     & $\ga$ 15.58  & $\la$ 1.70          & $\ga$ 15.58 & $\la$ 1.70         \\
    \hline
 \end{tabular}
 \tablenotetext{a}{numbers in brackets are 1$\sigma$ uncertainties referring}
 \tablenotetext{ }{to the last significant digits quoted}
 \tablenotetext{b}{$J$- and $H$-band data are from \citet{shibanov2003AsAp}}
 \tablenotetext{c}{mid-IR data are from \citet{danilenko2011MNRAS}}
 \label{t:flux}
 \end{center}
\end{table*}
\subsection{Photometry}
For stellar-like object 
photometry we used  a circular aperture  
with four pixels radius, 
comparable  to the effective seeing  
of the $K_s$ image. 
The correction for finite aperture, derived from the point spread function (PSF) 
of bright stars located close to the pulsar, is $\delta m$~= 0.64(2). 
In this case, uncertainties include the PSF 
variation over the selected image section.
For  the counter-jet  we used a polygon aperture  shown in Fig.~\ref{fig:gsao-im}, although  
the final results  depend weakly  on   
the specific aperture shape and 
background region parameters.
 
We have also remeasured the spatially integrated  brightnesses 
of the counter-jet in the $J_s$ and $H$ bands. 
The photometry was performed 
on the VLT/ISAAC pulsar subtracted images \citep[Fig.~1]{shibanov2003AsAp}. 
Our results differ significantly from those presented by \citet{danilenko2011MNRAS}.  
The reason is that the authors used mean surface brightnesses 
from \citet{shibanov2003AsAp} which were  accidentally 
overestimated by  1.94 $mag~arcsec^{-2}$ in both bands. 
After this correction the fluxes from \citet{shibanov2003AsAp}  are consistent with 
our ones. All measured fluxes were de-reddened 
using $E_{B-V}$~= 0.055(5) and $R_V$~= 3.1 \citep{shibanov2003AsAp}.
The results are summarized in Table~\ref{t:flux}.

For completeness, $K_s$ magnitudes of o1, o2, o3, o4 field objects 
are 22\fm48(8), 20\fm53(4), 18\fm77(4), and 23\fm2(2), 
respectively. The derived 3$\sigma$ detection limit for a point-like object 
for a 0\farcs3 aperture (corresponding to $\ga$~90\% of a point-like source flux) 
centered at some position in the pulsar vicinity free
from any sources is 23\fm6.

\subsection{Spectra of the Vela pulsar and its 
"counter-jet"}
\begin{figure}[t]
  \epsscale{1}
  \plotone{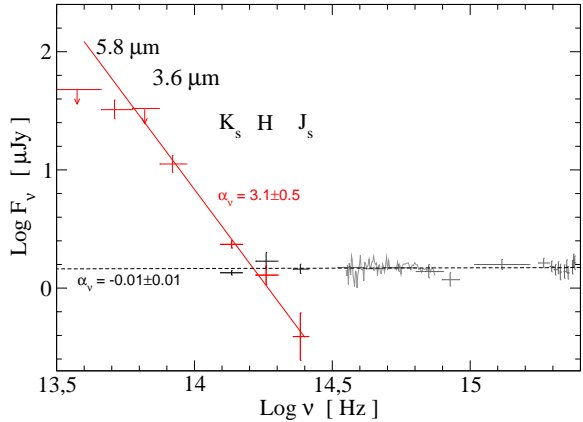}
  \caption{IR-UV spectra of the Vela pulsar (black) and 
  the likely counter-jet feature (red). 
  The UV-optical data are from \citet{romani2005ApJ} and \citet{mignani2007AsAp}, the IR data 
  are from \citet{shibanov2003AsAp} and  \citet{danilenko2011MNRAS}. 
  The dashed line approximates the flat UV--optical spectrum of the pulsar. The red line is the best fit 
  to the near-IR fluxes of the counter-jet.  
  The index $\alpha_{\nu}$ is defined as $F_{\nu}\propto \nu^{-\alpha_{\nu}}$.
  }
  \label{fig:multi_spec}
\end{figure}

Using our photometry results, in Fig.~\ref{fig:multi_spec} we  
show an upgrade of the IR-UV spectra of the pulsar and its likely 
counter-jet  compiled recently  by 
\citet{danilenko2011MNRAS}. 
One sees that
the $K_s$ flux of the pulsar agrees well
with the extrapolation of its flat  power law UV-optical  spectrum toward the longer wavelengths. 
This makes the Vela pulsar spectral energy distribution (SED) 
similar to that of the younger Crab  \citep{sandberg2009AsAp}. 

At the same time, the counter-jet spatially integrated $K_s$ 
flux (shown by red symbol in Fig.~\ref{fig:multi_spec})
is about twice as high as the pulsar flux. 
The counter-jet $J_sHK_s$ fluxes demonstrate 
a very steep power law SED  
 with a spectral index of about three. 
The Spitzer fluxes are compatible with the long wavelength extrapolation of this SED, suggesting a common  
nature of the near-IR and mid-IR emission. We have checked that 
the essentially non-thermal, likely synchrotron, spectrum of the counter-jet is different from
blackbody-like SEDs of the nearby stellar objects o1, o2, o3. This means that 
the extended feature can hardly be just a combination 
of faint blended stars. 

\section{Discussion}
The Gemini-South ground-based observations with the 
new generation of the AO system 
have provided us   
with the superb  image quality  comparable to that of the 
\textit{HST}. This   
allowed us to firmly detect, for the first time,
the Vela pulsar in the $K_s$ band. 
The measured  pulsar flux is consistent with the extrapolation of the pulsar optical-UV spectrum towards the IR, 
which shows that the spectrum remains flat in this range and does not demonstrate any excess suggested 
early by  the lower spatial resolution 
IR data. The AO observations  also enabled us to confirm and  
resolve the feature extended immediately behind the pulsar.   
The elongation  of the extended feature in the direction of  the X-ray counter-jet 
and in the opposite direction of the pulsar proper motion suggests that it can be 
associated with the X-ray counter-jet. 
This is supported by the compact and relatively bright knot-like structure 
within the feature (Fig.~\ref{fig:gsao-im}), which is reminiscent of the well known 
optical-near-IR knot within the south-east jet of the Crab pulsar.

The likely non-thermal spectrum of the Vela IR counter-jet is also similar 
to the  spectrum of the Crab knot, which follows the power law  with a 
spectral index of 1 and is responsible for 
an apparent excess of the Crab-pulsar mid-IR fluxes observed with
\textit{Spitzer} \citep{sandberg2009AsAp}.
The spectrum of the Vela counter-jet is even steeper resulting in a 
much stronger apparent excess of the Vela flux in the mid-IR.
Our data practically rule out alternative interpretations of the mid-IR excess    
discussed by \citet{danilenko2011MNRAS}, such as the fall-back disk or  
the complicated emitting particle distribution function in the pulsar magnetosphere.  
The Vela counter-jet spectrum is also  consistent  with typically red  
spatially integrated optical-IR spectra  of PWNe \citep{zharikov2013AsAp}.

The Crab knot \citep{sandberg2009AsAp}, and Vela jets \citep{pavlov2001ApJb} are  known to demonstrate 
a high temporal variability even on a week scale.  
The current data do not allow one to infer whether the Vela IR counter-jet is variable and move  
together with the pulsar. 
Detection of such variability would thus serve as a strong evidence for its PWN  nature.   
This can also help to discard possible alternative interpretations, such as a Vela supernova remnant filament 
or a background galaxy at a cosmological distance.   
Some marginal evidence of the feature variability has 
been reported by  \citet{shibanov2003AsAp} based on the VLT near-IR observations. 
Forthcoming VLT/NACO and \textit{Spitzer} data will hopefully find out whether the detected feature is variable,   
clarify its spectrum and  real nature. Observations at longer wavelengths are important 
to confirm power law spectrum of the feature.  
To detect other parts of the Vela PWN, deeper near-IR observations are
necessary.     

\acknowledgments

We are grateful to the Gemini AO system team for performing excellent observations and to the anonymous 
referee for useful comments. 
The work was partially supported 
by the Russian Foundation for Basic
Research (grants 11-02-00253 and 13-02-12017-ofi-m), RF Presidential Program (Grant NSh 4035.2012.2), 
and by CONACYT 151858 project.
REM acknowledges support by the BASAL 
Centro de Astrofisica y Tecnologias Afines (CATA) PFB--06/2007.



{\it Facilities:} \facility{Gemini}.  

\bibliographystyle{apj}
\bibliography{vela002}

\clearpage

\clearpage


\end{document}